\documentclass[%
prl,
 reprint,
 amsmath,amssymb,
 aps,preprintnumbers]{revtex4-1}

\usepackage{amssymb,amsmath}
\usepackage{textcomp}
\usepackage{tikz}
\usepackage{graphicx}
\usepackage{float}
\usepackage{textpos}
\usepackage{mathtools}
\usepackage{xcolor}
\usepackage{verbatim}

\usepackage{graphicx} 

\def\XXint#1#2#3{{\setbox0=\hbox{$#1{#2#3}{\int}$}
     \vcenter{\hbox{$#2#3$}}\kern-.5\wd0}}

\usepackage[colorlinks=true,linkcolor=black, citecolor=black,
urlcolor=black]{hyperref}
\usepackage{xcolor}

\usepackage{multirow,graphics}
    \newcommand{\beq}{\begin{equation}}
    \newcommand{\eeq}{\end{equation}}
    \newcommand\beqa{\begin{eqnarray}}
    \newcommand\eeqa{\end{eqnarray}}

\renewcommand{\geq}{\geqslant}

\usepackage{amstext}
\usepackage{amssymb}
\usepackage{amsmath}
\usepackage{graphicx}
\usepackage{color}

\usepackage{textcomp}
\usepackage{tikz}
\usepackage{graphicx}
\usepackage{float}

\usepackage{xcolor}

\usepackage{esint}

\usepackage{bbm}

\newcommand{\h}{ {\mathbbm{h}}}
\newcommand{\B}{ {\mathbbm{B}}}

\begin{document}

\makeatletter
     \@ifundefined{usebibtex}{\newcommand{\ifbibtexelse}[2]{#2}} {\newcommand{\ifbibtexelse}[2]{#1}}
\makeatother

\preprint{}

\newcommand{\footnoteab}[2]{\ifbibtexelse{%
\footnotetext{#1}%
\footnotetext{#2}%
\cite{Note1,Note2}%
}{%
\newcommand{\textfootnotea}{#1}%
\newcommand{\textfootnoteab}{#2}%
\cite{thefootnotea,thefootnoteab}}}

\title{
 Large-$N$ Principal Chiral Model in Arbitrary External Fields}

\author{Vladimir Kazakov}
\email{kazakov $\bullet$ lpt.ens.fr}

\affiliation{ Laboratoire de physique de l'\'Ecole normale sup\'erieure, CNRS,
Universit\'{e} PSL, Sorbonne Universit\'es,  24 rue Lhomond, 75005 Paris, France}

\author{Evgeny Sobko}
\email{evgenysobko $\bullet$ gmail.com}
\affiliation{London Institute for Mathematical Sciences, Royal Institution, London, W1S 4BS, UK}
        
\affiliation{ Laboratoire de physique de l'\'Ecole normale sup\'erieure, CNRS,
Universit\'{e} PSL, Sorbonne Universit\'es,  24 rue Lhomond, 75005 Paris, France}

\author{ Konstantin Zarembo}
\email{zarembo $\bullet$ nordita.org}
\affiliation{Nordita, KTH Royal Institute of Technology and Stockholm University, Stockholm, Sweden
}

\affiliation{Niels Bohr Institute, Copenhagen University, Copenhagen, Denmark
}

\begin{abstract}
We report the explicit solution for the vacuum state of the two-dimensional \(SU(N)\) Principal Chiral Model at large-\(N\) for an arbitrary set of chemical potentials and any interaction strength, a unique result of such  kind for an asymptotically free QFT. The solution matches
one-loop perturbative calculation at weak coupling, and in the opposite strong-coupling regime exhibits an emergent spacial dimension from the continuum limit of the $SU(N)$ Dynkin diagram.
\end{abstract}

\maketitle
\section{Introduction}

The Principal Chiral  Model (PCM) is a remarkable representative of a large class of integrable relativistic $1+1$-dimensional QFTs~\cite{Polyakov:1984et,Wiegmann:1984ec} possessing key features of QCD: asymptotic freedom, mass-gap generation and a nontrivial topological expansion in ’t Hooft’s large-\(N\) limit. These features have attracted considerable attention to the study of non-perturbative dynamics of PCM, via the exact $S$-matrix~\cite{Zamolodchikov:1978xm,Berg:1977dp,Wiegmann:1984ec}, and Bethe ansatz~\cite{Wiegmann:1984mk,Wiegmann:1984pw,Balog:1992cm,Balog:2003yr,Gromov:2008gj,Leurent:2015wzw} giving access to thermodynamics of the model~\cite{Wiegmann:1984ec}. While the $S$-matrix is explicitly known for PCM~\cite{Wiegmann:1984mk},  the linear Bethe equations for the vacuum as a function of chemical potentials or the non-linear TBA equations are difficult to solve at any coupling strength. They are usually studied either perturbatively~\cite{Balog:1992cm} or numerically~\cite{Balog:2003yr,Gromov:2008gj,Leurent:2015wzw}.  Fortunately, there is an exception: the large-$N$ limit of the vacuum Bethe equations is exactly solvable~\cite{Fateev:1994ai,Fateev:1994dp}
for a specific choice of external fields  (chemical potentials of conserved charges) repeating the mass spectrum profile in their dependence on the $SU(N)$ Dynkin label \footnote{see also \cite{DiPietro:2021yxb} for another profile of chemical potentials admiting exact solution in the large-\(N\) limit}. In this work we lift this restriction and solve the model at large $N$ for any set of chemical potentials. It gives us the first exact analytic solution for an asymptotically free theory at general chemical potentials.

The FKW solution~\cite{Fateev:1994ai,Fateev:1994dp} obeys the semi-circle law in rapidity and can be systematically extended to higher orders in $1/N$, moreover a double-scling (DS) limit combining large-$N$ with strong coupling resums all orders of the $1/N$ expansion~\cite{Kazakov:2019laa}. In this regime the Dynkin labels give rise to a new emergent dimension in addition to the $1+1$ physical dimensions of space-time, as was pointed out already in~\cite{Fateev:1994ai,Fateev:1994dp}. The DS limit presumably describes a three-dimensional non-critical string theory, by analogy to the $c=1$ matrix quantum mechanics~\cite{Kazakov:1988ch} whose DS limit~\cite{Brezin:1990rb,Gross:1989vs,Parisi:1989dka,Ginsparg:1990as} is dual to non-critical strings in two dimensions~\cite{Das:1990kaa,Polchinski:1991uq,Boulatov:1991xz,Gross:1990md,Gross:1990ub,Kazakov:2000pm,Alexandrov:2002fh}.   

We will report the solution for the vacuum state of large-$N$ PCM with arbitrary chemical potentials. The rapidity dependence happens to be always semi-circular with support $(-B_a,B_a)$ varying along the Dynkin diagram. We derived an integral equation for the limit shape $B(\alpha )$ assuming its continuity. We then match the free energy with the known one-loop result~\cite{Balog:1992cm} at weak coupling. At strong coupling, we observe the locality along the third emergent dimension for the leading order of small deviations of external field from the first, Perron-Frobenius (PF) mode. This may be relevant for the physical properties of the effective theory of the dual three-dimensional string. 
 
\section{PCM with General Chemical Potentials at Large \(N\)}

The PCM  is defined by the Lagrangian 
\begin{gather}
S=\frac{N}{\lambda_0}\int d^2x\, \mathop{\mathrm{tr}} D_\mu g^\dag D^\mu g,
\end{gather}
where the field $g(x)\in SU(N)$. The mass spectrum consists of \(N-1\) particles transforming in the bi-fundamental representations of $SU(N)\times SU(N)$. 
The lightest particle acquires its mass by the mechanism of dimensional transmutation while others can be seen as bound states with masses :
\begin{gather}
m_a=m\frac{\sin\frac{\pi a}{N}}{\sin\frac{\pi}{N}}\,,\quad a=1\ldots N-1.
\end{gather} 
A finite density of particles is introduced by  gauging the $SU(N)\times SU(N)$ global symmetry by constant chemical potentials: $D_0=\partial_0 g-\frac{i}{2}(Hg+gH)$, $D_1=\partial_1$, where 
$H=\mathop{\mathrm{diag}}(q _1,\ldots ,q_N)$ is a traceless diagonal matrix with the eigenvalues organized in the descending order: $q _1\geq q _2\geq\ldots \geq q _N$. The chemical potential of the $a$-th species, transforming in the rank-$a$ anti-symmetric representation, is then $h_a=\sum_{b=1}^{a}q _b$. At large-$N$ we introduce a continuous coordinate $\alpha =\pi a/N\in(0,\pi )$ and define the limit shape functions $q(\alpha )=q_a$, $h(\alpha )=\pi h_a/N$,  smooth at $N\rightarrow\infty$ and related by $h'(\alpha )=q (\alpha )$. Trace and ordering conditions on $q_a$ translate into the boundary conditions on $h(\alpha )$:  
\begin{equation}
 h(0)=0=h(\pi ),\qquad h''(\alpha )<0.
\end{equation}
 
The Bethe equations for the ground state of PCM take the form of integral equations for the pseudo-energies~\cite{Fateev:1994ai}: 
\begin{equation}
\varepsilon_a(\theta)+\sum\limits_{b=1}^{N-1}\int\limits_{-B_b}^{B_b}d\theta \,{R}_{ab}(\theta-\theta')\varepsilon_b(\theta)=h_a-m_a\cosh \theta,\label{InitialEqnsWithDelta}
\end{equation}
which hold on the intervals $(-B_a,B_a)$ where the pseudo-energies are positive. At the endpoints they vanish: $\varepsilon _a(\pm B_a)=0$, which is an extra condition closing the system.
Once the pseudo-energies are known, the vacuum energy is obtained by simple integration:
\begin{equation}\label{originalE}
E=-\sum\limits_{a=1}^{N-1}m_a\int\limits_{-B_a}^{B_a}\frac{d\theta}{2\pi }\,\, \varepsilon_a(\theta)\cosh \theta.
\end{equation}

The kernels of the integral equation originate from scattering between particles in the Fermi sea and can be extracted from the exact S-matrix of PCM. The kernel (its Fourier transform  is explicitly given  in~\cite{Fateev:1994ai}) is
\begin{equation}\label{kernel}
 R_{ab}(\theta )=G\left(\frac{\pi (a+b)}{N}+i\theta \right)
 -G\left(\frac{\pi |a-b|}{N}+i\theta \right)+{\rm c.c.}
\end{equation}
where
\begin{equation}
 4\pi ^2G(x)=\psi \left(\frac{x}{2\pi }\right)+\psi \left(-\frac{x}{2\pi }\right)
 -\frac{2\pi }{x}\,,
\end{equation}
and $\psi (x)$ is the logarithmic derivative of the gamma-function.

\begin{figure}
   \includegraphics[scale=0.34]{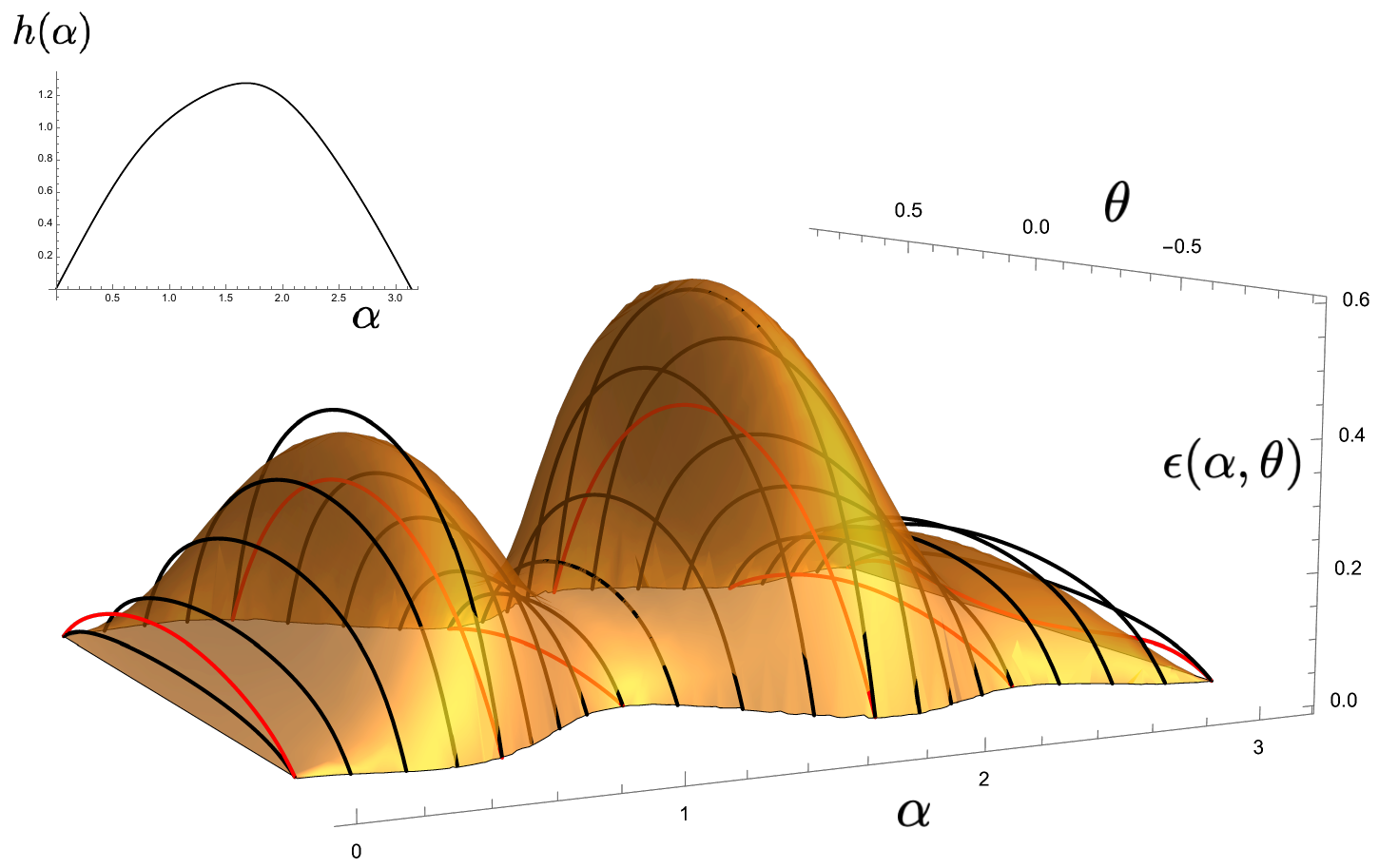}
   \label{fig:Ng1} 
\caption{The structure of PCM vacuum for an example of a general profile of chemical potentials. Semitransparent surface represents pseudo-energy \(\epsilon(\alpha,\theta)\) \eqref{ContinuousSemiCircle} drawn from the numerical solution of equations \eqref{brief-shape} for a given limit profile of supports \(B(\alpha)\). The numerical solution of the original Bethe equations \eqref{InitialEqnsWithDelta} for \(N=7\) and \(N=22\) are drawn by the red and black lines, correspondingly, demonstrating the convergence to the large \(N\) solution. The corresponding profile of the chemical potentials is drawn at the upper-left corner.}
\end{figure}

Taking the large-$N$ limit is straightforward: the Dynkin labels are promoted to continuous variables, summation is replaced by integration, and the pseudo-energies become functions of two variables $\alpha$ and $\theta$. Quite remarkably, the Dynkin coordinate and rapidity naturally combine into complex combinations:
\begin{equation}
 z=\alpha +i\theta ,\qquad w=\beta +i\theta ',
\end{equation}
such that the kernel (\ref{kernel})  becomes almost a sum of holomorphic and antiholomorphic functions:
\begin{eqnarray}
R(z,\bar{z};w,\bar{w})&=&G(z+\bar{w})-\frac{1}{2}\,G(z-w)-\frac{1}{2}\,G(w-z)
\nonumber \\
  &&+ \frac{\mathop{\mathrm{sign}}(\alpha -\beta )}{2\pi (z-w)}+{\rm c.c.}
\end{eqnarray}
The anomalous, holomorphicity-breaking term contains the regularized $\mathop{\mathrm{sign}}$ function, that equals $\pm 1$ outside the interval $(-\varepsilon ,\varepsilon )$ and zero inside. Here $\varepsilon $ is a proxy for $1/N$, but its precise numerical value will not be important since at the end it just defines the contour deformation prescription.
Finally we get the Bethe equations in the form of two-dimensional integral equation :
\begin{equation}\label{int-hol}
 \int\limits_{\mathcal{D}}^{}d^2w\,R(z,\bar{z};w,\bar{w})\varepsilon (w,\bar{w})
 =h(\alpha )-\frac{m}{2}(\sin z+\sin\bar{z}),
\end{equation}
where integration is over a complex domain 
\begin{equation}
 \mathcal{D}=\{\alpha +i\theta \,|\,0< \alpha <\pi ,\,-B(\alpha )<\theta< B(\alpha ) \}.
\end{equation}
The equation  holds only inside this domain.

\section{Solution of Bethe equation}

Solution of the large-$N$ integral equation crucially relies on holomorphy of the kernel. Every holomorphic function is harmonic and thus applying the Laplacian $\partial ^2/\partial z\partial \bar{z}$ to both sides of (\ref{int-hol})  picks only the anomalous term which moreover becomes local in the Dynkin variable. At the end we get a singular integral equation for each $\alpha $:
\begin{gather}\label{sing-int}
\frac{2}{\pi }\fint\limits_{-B(\alpha)}^{B(\alpha)}d\theta'\,\,\frac{\varepsilon(\alpha,\theta')}{(\theta-\theta')^2}=h''(\alpha) ,
\end{gather}
the solution of which is the famous semi-circle law:
\begin{gather}\label{ContinuousSemiCircle}
\varepsilon(\alpha,\theta)=-\frac{h''(\alpha)}{2}\sqrt{B(\alpha)^2-\theta^2}\,.
\end{gather}

Another line of argument starts with a function
\begin{eqnarray}\label{def-F}
 F(z)&=&\int\limits_{\mathcal{D}}^{}
 d^2w\,\left[
 \vphantom{\frac{\mathop{\mathrm{sign}}(\alpha -\beta )}{\pi (z-w)}}
 2G(z+\bar{w})-G(z-w)-G(w-z)
 \right.
\nonumber \\
 &&\left.
 +\frac{\mathop{\mathrm{sign}}(\alpha -\beta )}{\pi (z-w)}
 \right]\varepsilon (w,\bar{w})-{i}h'(\alpha )\theta -h(\alpha ).
\end{eqnarray}
Holomorphicity of this function in $\mathcal{D}$ is precisely equivalent to the integral equation (\ref{sing-int}), which can be checked by computing $\partial F /\partial \bar{z}$. On the other hand,
\begin{equation}
  \int\limits_{\mathcal{D}}^{}d^2w\,R(z,\bar{z};w,\bar{w})\varepsilon (w,\bar{w})
 =\frac{F(z)}{2}+\frac{F(\bar{z})}{2}+h(\alpha ).
\end{equation}
Provided (\ref{sing-int}) is satisfied, the integral equation (\ref{int-hol}) factorizes into holomorphic and anti-holomorphic parts and boils down to a functional equation in one variable:
\begin{equation}\label{shape1}
 F(z)=-m\sin z.
\end{equation}
This equation does not follow from (\ref{sing-int}), it is an extra condition that can be used for fixing the profile of the Fermi surface $B(\alpha )$. We thus call it the {\it shape equation}.

A useful form of the shape equation is obtained by 
turning the branch cuts of the square root inside out, in other words replacing the short cut  $[-B(\alpha ), B(\alpha )]$ passing through zero by a long cut passing through infinity. The resulting expression is manifestly analytic in $\mathcal{D}$:
\begin{eqnarray}\label{integratedF}
 F(z)&=&\int\limits_{0}^{\pi }\frac{d\beta }{2\pi i}\,\,h''(\beta )\sin\beta 
 \int\limits_{B(\beta )}^{\infty } \frac{d\xi\,\sqrt{\xi ^2-B(\beta )^2}}{\cosh(\xi -iz)+\cos\beta } 
\nonumber \\
 &&-(z\rightarrow -z).
 \end{eqnarray}
We mention in passing that transformation from short to long cuts is well familiar in the theory of integrable systems, for example, it is a crucial step in deriving the Quantum Spectral Curve for AdS/CFT \cite{Gromov:2013pga,Gromov:2014caa}. Apart from making holomorphy manifest, this representation reveals the symmetries of the problem:  $F(z)$ is evidently $2\pi $-periodic and anti-symmetric. A convenient basis in the space of such functions is $\{\sin pz\,|\,p\in\mathbbm{N}\}$. Expanding (\ref{shape1}), (\ref{integratedF}) in this basis and matching the coefficients gives:
\begin{equation}\label{brief-shape}
 \int\limits_{0}^{\pi }d\alpha \,h''(\alpha ) B(\alpha )K_1(pB(\alpha ))\sin p\alpha=-\frac{\pi m}{2}\,\delta _{p1},
\end{equation}
where $K_1$ is the modified Bessel function. For the energy we get according to (\ref{originalE}):
\begin{equation}
 E=\frac{N^2m}{4\pi ^2}\int\limits_{0}^{\pi }d\alpha \,h''(\alpha )
 B(\alpha )I_1(B(\alpha ))\sin \alpha.
\end{equation}
These two equations completely characterize the ground state at arbitrary chemical potentials. 

If the chemical potentials repeat the mass spectrum, with only the lowest PF mode present: $h(\alpha )=\h\sin\alpha $, the solution is constant: $B(\alpha )=\B$. Indeed, the shape equations  (\ref{brief-shape})  with $p\geq 2$ then follow from orthogonality of $\sin k\alpha $ with different $k$, and the shape equation for $p=1$ gives
\begin{equation}\label{old-sol}
 \B K_1(\B)=\frac{m}{\h}\,,\qquad E=-\frac{N^2m\h}{8\pi }\,\B I_1(\B),
\end{equation}
recovering the results of~\cite{Fateev:1994ai,Fateev:1994dp}. At large chemical potential the theory is weakly coupled, and the Fermi rapidity grows logarithmically:
\begin{equation}
 \B\stackrel{\h\gg m}{\simeq } \ln\frac{\h}{m}+\frac{1}{2}\,\ln\ln\frac{\h}{m}+\frac{1}{2}\ln\frac{\pi}{2}
 +\ldots 
\end{equation}
and satisfies the two-loop RG equation. It can thus
 be identified with the inverse of the running coupling in a particular renormalization scheme~\cite{Fateev:1994ai}:
\begin{equation}\label{running-lambda}
 \frac{4\pi }{\lambda (\h)}=\B.
\end{equation}
This identification gives a useful qualitative picture of the Bethe Ansatz solution where the Fermi rapidity directly controls the interaction strength. 
In the strong-coupling regime (small $\B$), we have~\cite{Fateev:1994ai}:
\begin{equation}\label{small-B}
 \B\stackrel{\h\rightarrow m}{\simeq }\sqrt{\frac{4\Delta }{|\ln\Delta |}}\,,\qquad 
 \Delta =\frac{\h}{m}-1\,.
\end{equation}

In general, the shape equation is an infinite set of non-linear integral conditions for a real function $B(\alpha )$. Solving them is difficult and here we will restrict to small pertubations around the constant solution. 

\section{Small fluctuations around PF mode}

Consider the chemical potential of the form
\begin{equation}\label{our-h}
 h(\alpha )=\h\sin\alpha +\sum_{p=2}^{\infty }h_p\sin p\alpha ,\qquad h_p\ll \h.
\end{equation}
The Fermi surface will fluctuate around the constant: $B(\alpha )=\B+b(\alpha )$,  and we may expand the shape equations in $b$. To streamline perturbation theory we first introduce some convenient notations.

We define a pairing:
\begin{equation}
 \left\langle f\right.\!\left|g \right\rangle
 =-\frac{2}{\pi }\int\limits_{0}^{\pi }d\alpha \,h''(\alpha )f(\alpha )^*g(\alpha ),
\end{equation}
and two sets of functions:
\begin{equation}
 \left\langle p\right|=\sin p\alpha ,\qquad \left|p\right\rangle=U_{p-1}(\cos\alpha ),
\end{equation}
where $U_n$ are Chebyshev polynomials. Those are orthogonal for the pure PF profile, while for (\ref{our-h}):
\begin{equation}
 \left\langle p\right.\!\left|p' \right\rangle=\h\delta _{pp'}+\mathcal{O}(h_k).
\end{equation}
The $\mathcal{O}(h_k)$ terms can be also computed,  we will only need 
\begin{equation}
 \left\langle 1\right.\!\left|p \right\rangle=\left\langle p\right.\!\left| 1 \right\rangle =p^2h_p,\qquad \left\langle 1\right.\!\left| 1\right\rangle
 =\h.
\end{equation}

In these notations, the shape equations and the energy are concisely written as
\begin{equation}
 \left\langle p\right.\!\left|\mathcal{K}(pB) \right\rangle=m\delta _{p1},
 \qquad 
 E=-\frac{N^2m}{8\pi }\,\left\langle 1\right.\!\left| \mathcal{I}(B)\right\rangle,
\end{equation}
where $\mathcal{K}(x)=xK_1(x)$ and $\mathcal{I}(x)=xI_1(x)$.

It is now straightforward to expand the shape equations around the constant solution:
\begin{equation}
 \mathcal{K}\left\langle p\right.\!\left|1 \right\rangle
 +p\mathcal{K}'\left\langle p\right.\!\left|b \right\rangle
 +\frac{1}{2}\,p^2\mathcal{K}''\left\langle p\right.\!\left|b^2 \right\rangle
 +\ldots =m\delta _{p1}.
\end{equation}
If $p\geq 2$, we may keep only the linear order:
\begin{equation}
 \left\langle p\right.\!\left|b \right\rangle=-\frac{p\mathcal{K}}{\mathcal{K}'}\,h_p,
\end{equation}
from which 
\begin{equation}\label{b-sol}
 b(\alpha )=\sum_{p=2}^{\infty }\frac{pK_1(p\B)}{\h K_0(p\B)}\,h_pU_{p-1}(\cos\alpha ).
\end{equation}
The constant $p=1$ mode in $b(\alpha )$ appears at the quadratic order and can be found from the $p=1$ equation because the linear term there cancels idetically:
\begin{equation}\label{b^2-sol}
 \left\langle 1\right.\!\left| b\right\rangle=-\frac{\mathcal{K''}}{2\mathcal{K}'}\,\left\langle 1\right.\!\left| b^2\right\rangle,
\end{equation}
where for $b$ on the right-hand side we can use the linear approximation above. 

Corrections to the energy appear to be quadratic:
\begin{equation}
 E= -\frac{N^2m}{8\pi }\left(\mathcal{I}\left\langle 1\right.\!\left|1 \right\rangle
 +\mathcal{I}'\left\langle 1\right.\!\left|b \right\rangle
 +\frac{1}{2}\,\mathcal{I}''\left\langle 1\right.\!\left| b^2\right\rangle
+\ldots  \right).
\end{equation}
Using (\ref{b^2-sol}) and plugging in the solution for $b$ we get:
\begin{equation}
 E=-\frac{N^2m}{8\pi \h}\left(
 \B I_1(\B)\h^2
 +\sum_{p=2}^{\infty }
 \frac{p^2K_1(p\B)^2}{2K_0(\B)K_0(p\B)^2}\,h_p^2
 \right).
\end{equation}
This result is valid both at weak and strong coupling, the only assumption used in its derivation is smallness of deviations from the PF mode in $h(\alpha)$.

\paragraph{Weak coupling.}
At weak coupling (large $\B$), we have:
\begin{equation}\label{Eweak}
 E\stackrel{\B\gg 1}{\simeq }-\frac{N^2}{16\pi }
 \left[
 \B\h^2+\B\sum_{p}^{}h_p^2
 +\frac{1}{2}\,\sum_{p}^{}p(p+2)h_p^2
 \right].
\end{equation}
This can be compared with the perturbative expression know to the one-loop order ~\cite{Balog:1992cm} for generic chemical potentials~\footnote{The original expression in \cite{Balog:1992cm} was computed in the $\overline{MS}$ scheme. A finite renormalization constant connecting $\overline{MS}$ to the scheme defined by (\ref{running-lambda}) was computed in \cite{Fateev:1994ai}, and has been already taken into account in (\ref{E1-loop}).}:
\begin{eqnarray}\label{E1-loop}
 &&E^{{\rm 1-loop}}=-\frac{N^2}{2\pi \lambda (\h)}\int_{}^{}d\alpha \,q(\alpha )^2
\nonumber \\
&&
 -\frac{N^2}{16\pi ^3}\int_{}^{}d\alpha\, d\beta \,\left(q(\alpha )-q(\beta )\right)^2
 \ln\frac{2\left|q(\alpha )-q(\beta )\right|}{{\rm e}\h}
\end{eqnarray}
where $q(\alpha )=h'(\alpha )$ and the running coupling is defined by (\ref{running-lambda}). Plugging in $h(\alpha )$ from (\ref{our-h}) and expanding in $h_p$ to the quadratic order we get the complete agreement with our integrability-based result.

\paragraph{Strong coupling.} In the strong-coupling regime $\h$ approaches $m$ and $\B$ goes to zero, then
\begin{equation}
 E\stackrel{\B\ll 1}{\simeq }-\frac{N^2}{16\pi }\left(
 \B^2\h^2+\frac{1}{\B^2|\ln\B|^3}\,\sum_{p}^{}h_p^2
 \right).
\end{equation}
Quite remarkably, the energy becomes local along the Dynkin dimension. Using (\ref{small-B}), and denoting $\varphi(\alpha )=(h(\alpha )-\h\sin\alpha )/(4\Delta) $, we get:
\begin{equation}
 E\stackrel{\Delta \rightarrow 0}{\simeq }-\frac{N^2\Delta }{4\pi |\ln\Delta| }\left(m^2
 +\frac{1}{|\ln\Delta| }\,\int_{}^{}d\alpha \,\varphi (\alpha )^2
 \right).
\end{equation}

Deviations from the PF mode are naturally of order $\Delta $, thus $\varphi $ can be viewed as an order-one variable, albeit we had linearized in fluctuations and thus neglected contributions of higher than quadratic order in $\varphi $, moreover one can show that the very leading term will be the same for any small profile \(B(\alpha)\).  The fluctuations beyond the PF mode appear to be logarithmically suppressed at strong coupling. Perhaps the PF contribution should be regarded as a subtraction constant and the remainder could possibly be interpreted in terms of an effective theory for fluctuations, of which we calculated but the quadratic order.

\section{Discussion}

We extended the exact solution of large-$N$ PCM ~\cite{Fateev:1994ai,Fateev:1994dp} to the case of arbitrary chemical potential. The pseudo-energies of physical excitations always  follow the semi-circular law but now the Fermi surface becomes a new functional degree of freedom that has a non-trivial profile along the Dynkin diagram. This plays a role of a new continuous dimension emergent in the large-$N$ limit. With such new functional degree of freedom at hands one can model various weak coupling regimes which may reveal new types of  nonperturbative trans-series expansions \cite{Marino:2019eym}.
At strong coupling, intriguing signs of locality  arise, perhaps pointing towards string theory interpretation.  It would be interesting to split the Dynkin diagram into two intervals and calculate entanglement entropy as a diagnostic for locality \cite{Hartnoll:2015fca}. 

Similarly to our previous work \cite{Kazakov:2019laa} one can identify the Double Scaling limit in the case of general chemical potentials, we will return to this point in the forthcoming publication \cite{KSZ_Long}. 

Another interesting direction would be to analyze the large-\(N\) limit of PCM compactified on the cylinder with twisted boundary conditions, the corresponding analysis at finite \(N\) was already initiated in \cite{Leurent:2015wzw}. This corresponds to the putative dual string at finite temperature.
Also it would be interesting to go beyond the vacuum state and analyse the excited states.

\begin{acknowledgments}
\section*{Acknowledgments}
\label{sec:acknowledgments}

 V.K.
thanks Perimeter Institute where a part of this work was done, for hospitality.
This research was supported in part by Perimeter Institute for Theoretical Physics. Research
at Perimeter Institute is supported by the Government of Canada through the Department
of Innovation, Science, and Economic Development, and by the Province of Ontario through
the Ministry of Colleges and Universities.
The work of K.~Z. was supported by VR grant 2021-04578. 
Nordita is partially supported by Nordforsk.
\end{acknowledgments}

\bibliography{PCF_large_N.bib}

\begin{thebibliography}{33}%
\makeatletter
\providecommand \@ifxundefined [1]{%
 \@ifx{#1\undefined}
}%
\providecommand \@ifnum [1]{%
 \ifnum #1\expandafter \@firstoftwo
 \else \expandafter \@secondoftwo
 \fi
}%
\providecommand \@ifx [1]{%
 \ifx #1\expandafter \@firstoftwo
 \else \expandafter \@secondoftwo
 \fi
}%
\providecommand \natexlab [1]{#1}%
\providecommand \enquote  [1]{``#1''}%
\providecommand \bibnamefont  [1]{#1}%
\providecommand \bibfnamefont [1]{#1}%
\providecommand \citenamefont [1]{#1}%
\providecommand \href@noop [0]{\@secondoftwo}%
\providecommand \href [0]{\begingroup \@sanitize@url \@href}%
\providecommand \@href[1]{\@@startlink{#1}\@@href}%
\providecommand \@@href[1]{\endgroup#1\@@endlink}%
\providecommand \@sanitize@url [0]{\catcode `\\12\catcode `\$12\catcode `\&12\catcode `\#12\catcode `\^12\catcode `\_12\catcode `\%12\relax}%
\providecommand \@@startlink[1]{}%
\providecommand \@@endlink[0]{}%
\providecommand \url  [0]{\begingroup\@sanitize@url \@url }%
\providecommand \@url [1]{\endgroup\@href {#1}{\urlprefix }}%
\providecommand \urlprefix  [0]{URL }%
\providecommand \Eprint [0]{\href }%
\providecommand \doibase [0]{http://dx.doi.org/}%
\providecommand \selectlanguage [0]{\@gobble}%
\providecommand \bibinfo  [0]{\@secondoftwo}%
\providecommand \bibfield  [0]{\@secondoftwo}%
\providecommand \translation [1]{[#1]}%
\providecommand \BibitemOpen [0]{}%
\providecommand \bibitemStop [0]{}%
\providecommand \bibitemNoStop [0]{.\EOS\space}%
\providecommand \EOS [0]{\spacefactor3000\relax}%
\providecommand \BibitemShut  [1]{\csname bibitem#1\endcsname}%
\let\auto@bib@innerbib\@empty
\bibitem [{\citenamefont {Polyakov}\ and\ \citenamefont {Wiegmann}(1984)}]{Polyakov:1984et}%
  \BibitemOpen
  \bibfield  {author} {\bibinfo {author} {\bibfnamefont {A.~M.}\ \bibnamefont {Polyakov}}\ and\ \bibinfo {author} {\bibfnamefont {P.~B.}\ \bibnamefont {Wiegmann}},\ }\href {\doibase 10.1016/0370-2693(84)90206-5} {\bibfield  {journal} {\bibinfo  {journal} {Phys. Lett.}\ }\textbf {\bibinfo {volume} {141B}},\ \bibinfo {pages} {223} (\bibinfo {year} {1984})}\BibitemShut {NoStop}%
\bibitem [{\citenamefont {Wiegmann}(1984{\natexlab{a}})}]{Wiegmann:1984ec}%
  \BibitemOpen
  \bibfield  {author} {\bibinfo {author} {\bibfnamefont {P.}~\bibnamefont {Wiegmann}},\ }\href {\doibase 10.1016/0370-2693(84)91256-5} {\bibfield  {journal} {\bibinfo  {journal} {Phys. Lett.}\ }\textbf {\bibinfo {volume} {142B}},\ \bibinfo {pages} {173} (\bibinfo {year} {1984}{\natexlab{a}})}\BibitemShut {NoStop}%
\bibitem [{\citenamefont {Zamolodchikov}\ and\ \citenamefont {Zamolodchikov}(1979)}]{Zamolodchikov:1978xm}%
  \BibitemOpen
  \bibfield  {author} {\bibinfo {author} {\bibfnamefont {A.~B.}\ \bibnamefont {Zamolodchikov}}\ and\ \bibinfo {author} {\bibfnamefont {A.~B.}\ \bibnamefont {Zamolodchikov}},\ }\href {\doibase 10.1016/0003-4916(79)90391-9} {\bibfield  {journal} {\bibinfo  {journal} {Annals Phys.}\ }\textbf {\bibinfo {volume} {120}},\ \bibinfo {pages} {253} (\bibinfo {year} {1979})}\BibitemShut {NoStop}%
\bibitem [{\citenamefont {Berg}\ \emph {et~al.}(1978)\citenamefont {Berg}, \citenamefont {Karowski}, \citenamefont {Weisz},\ and\ \citenamefont {Kurak}}]{Berg:1977dp}%
  \BibitemOpen
  \bibfield  {author} {\bibinfo {author} {\bibfnamefont {B.}~\bibnamefont {Berg}}, \bibinfo {author} {\bibfnamefont {M.}~\bibnamefont {Karowski}}, \bibinfo {author} {\bibfnamefont {P.}~\bibnamefont {Weisz}}, \ and\ \bibinfo {author} {\bibfnamefont {V.}~\bibnamefont {Kurak}},\ }\href {\doibase 10.1016/0550-3213(78)90489-3} {\bibfield  {journal} {\bibinfo  {journal} {Nucl. Phys. B}\ }\textbf {\bibinfo {volume} {134}},\ \bibinfo {pages} {125} (\bibinfo {year} {1978})}\BibitemShut {NoStop}%
\bibitem [{\citenamefont {Wiegmann}(1984{\natexlab{b}})}]{Wiegmann:1984mk}%
  \BibitemOpen
  \bibfield  {author} {\bibinfo {author} {\bibfnamefont {P.~B.}\ \bibnamefont {Wiegmann}},\ }\href@noop {} {\bibfield  {journal} {\bibinfo  {journal} {JETP Lett.}\ }\textbf {\bibinfo {volume} {39}},\ \bibinfo {pages} {214} (\bibinfo {year} {1984}{\natexlab{b}})},\ \bibinfo {note} {[Pisma Zh. Eksp. Teor. Fiz.39,180(1984)]}\BibitemShut {NoStop}%
\bibitem [{\citenamefont {Wiegmann}(1984{\natexlab{c}})}]{Wiegmann:1984pw}%
  \BibitemOpen
  \bibfield  {author} {\bibinfo {author} {\bibfnamefont {P.~B.}\ \bibnamefont {Wiegmann}},\ }\href {\doibase 10.1016/0370-2693(84)90205-3} {\bibfield  {journal} {\bibinfo  {journal} {Phys. Lett.}\ }\textbf {\bibinfo {volume} {141B}},\ \bibinfo {pages} {217} (\bibinfo {year} {1984}{\natexlab{c}})}\BibitemShut {NoStop}%
\bibitem [{\citenamefont {Balog}\ \emph {et~al.}(1992)\citenamefont {Balog}, \citenamefont {Naik}, \citenamefont {Niedermayer},\ and\ \citenamefont {Weisz}}]{Balog:1992cm}%
  \BibitemOpen
  \bibfield  {author} {\bibinfo {author} {\bibfnamefont {J.}~\bibnamefont {Balog}}, \bibinfo {author} {\bibfnamefont {S.}~\bibnamefont {Naik}}, \bibinfo {author} {\bibfnamefont {F.}~\bibnamefont {Niedermayer}}, \ and\ \bibinfo {author} {\bibfnamefont {P.}~\bibnamefont {Weisz}},\ }\href {\doibase 10.1103/PhysRevLett.69.873} {\bibfield  {journal} {\bibinfo  {journal} {Phys. Rev. Lett.}\ }\textbf {\bibinfo {volume} {69}},\ \bibinfo {pages} {873} (\bibinfo {year} {1992})}\BibitemShut {NoStop}%
\bibitem [{\citenamefont {Balog}\ and\ \citenamefont {Hegedus}(2004)}]{Balog:2003yr}%
  \BibitemOpen
  \bibfield  {author} {\bibinfo {author} {\bibfnamefont {J.}~\bibnamefont {Balog}}\ and\ \bibinfo {author} {\bibfnamefont {A.}~\bibnamefont {Hegedus}},\ }\href {\doibase 10.1088/0305-4470/37/5/027} {\bibfield  {journal} {\bibinfo  {journal} {J. Phys.}\ }\textbf {\bibinfo {volume} {A37}},\ \bibinfo {pages} {1881} (\bibinfo {year} {2004})},\ \Eprint {http://arxiv.org/abs/hep-th/0309009} {arXiv:hep-th/0309009 [hep-th]} \BibitemShut {NoStop}%
\bibitem [{\citenamefont {Gromov}\ \emph {et~al.}(2009)\citenamefont {Gromov}, \citenamefont {Kazakov},\ and\ \citenamefont {Vieira}}]{Gromov:2008gj}%
  \BibitemOpen
  \bibfield  {author} {\bibinfo {author} {\bibfnamefont {N.}~\bibnamefont {Gromov}}, \bibinfo {author} {\bibfnamefont {V.}~\bibnamefont {Kazakov}}, \ and\ \bibinfo {author} {\bibfnamefont {P.}~\bibnamefont {Vieira}},\ }\href {\doibase 10.1088/1126-6708/2009/12/060} {\bibfield  {journal} {\bibinfo  {journal} {JHEP}\ }\textbf {\bibinfo {volume} {12}},\ \bibinfo {pages} {060} (\bibinfo {year} {2009})},\ \Eprint {http://arxiv.org/abs/0812.5091} {arXiv:0812.5091 [hep-th]} \BibitemShut {NoStop}%
\bibitem [{\citenamefont {Leurent}\ and\ \citenamefont {Sobko}(2015)}]{Leurent:2015wzw}%
  \BibitemOpen
  \bibfield  {author} {\bibinfo {author} {\bibfnamefont {S.}~\bibnamefont {Leurent}}\ and\ \bibinfo {author} {\bibfnamefont {E.}~\bibnamefont {Sobko}},\ }\href@noop {} {\  (\bibinfo {year} {2015})},\ \Eprint {http://arxiv.org/abs/1511.08491} {arXiv:1511.08491 [hep-th]} \BibitemShut {NoStop}%
\bibitem [{\citenamefont {Fateev}\ \emph {et~al.}(1994{\natexlab{a}})\citenamefont {Fateev}, \citenamefont {Kazakov},\ and\ \citenamefont {Wiegmann}}]{Fateev:1994ai}%
  \BibitemOpen
  \bibfield  {author} {\bibinfo {author} {\bibfnamefont {V.~A.}\ \bibnamefont {Fateev}}, \bibinfo {author} {\bibfnamefont {V.~A.}\ \bibnamefont {Kazakov}}, \ and\ \bibinfo {author} {\bibfnamefont {P.~B.}\ \bibnamefont {Wiegmann}},\ }\href {\doibase 10.1016/0550-3213(94)90405-7} {\bibfield  {journal} {\bibinfo  {journal} {Nucl. Phys.}\ }\textbf {\bibinfo {volume} {B424}},\ \bibinfo {pages} {505} (\bibinfo {year} {1994}{\natexlab{a}})},\ \Eprint {http://arxiv.org/abs/hep-th/9403099} {arXiv:hep-th/9403099 [hep-th]} \BibitemShut {NoStop}%
\bibitem [{\citenamefont {Fateev}\ \emph {et~al.}(1994{\natexlab{b}})\citenamefont {Fateev}, \citenamefont {Wiegmann},\ and\ \citenamefont {Kazakov}}]{Fateev:1994dp}%
  \BibitemOpen
  \bibfield  {author} {\bibinfo {author} {\bibfnamefont {V.~A.}\ \bibnamefont {Fateev}}, \bibinfo {author} {\bibfnamefont {P.~B.}\ \bibnamefont {Wiegmann}}, \ and\ \bibinfo {author} {\bibfnamefont {V.~A.}\ \bibnamefont {Kazakov}},\ }\href {\doibase 10.1103/PhysRevLett.73.1750} {\bibfield  {journal} {\bibinfo  {journal} {Phys. Rev. Lett.}\ }\textbf {\bibinfo {volume} {73}},\ \bibinfo {pages} {1750} (\bibinfo {year} {1994}{\natexlab{b}})}\BibitemShut {NoStop}%
\bibitem [{Note1()}]{Note1}%
  \BibitemOpen
  \bibinfo {note} {See also \cite {DiPietro:2021yxb} for another profile of chemical potentials admiting exact solution in the large-\(N\) limit}\BibitemShut {NoStop}%
\bibitem [{\citenamefont {Kazakov}\ \emph {et~al.}(2020)\citenamefont {Kazakov}, \citenamefont {Sobko},\ and\ \citenamefont {Zarembo}}]{Kazakov:2019laa}%
  \BibitemOpen
  \bibfield  {author} {\bibinfo {author} {\bibfnamefont {V.}~\bibnamefont {Kazakov}}, \bibinfo {author} {\bibfnamefont {E.}~\bibnamefont {Sobko}}, \ and\ \bibinfo {author} {\bibfnamefont {K.}~\bibnamefont {Zarembo}},\ }\href {\doibase 10.1103/PhysRevLett.124.191602} {\bibfield  {journal} {\bibinfo  {journal} {Phys. Rev. Lett.}\ }\textbf {\bibinfo {volume} {124}},\ \bibinfo {pages} {191602} (\bibinfo {year} {2020})}\BibitemShut {NoStop}%
\bibitem [{\citenamefont {Kazakov}\ and\ \citenamefont {Migdal}(1988)}]{Kazakov:1988ch}%
  \BibitemOpen
  \bibfield  {author} {\bibinfo {author} {\bibfnamefont {V.~A.}\ \bibnamefont {Kazakov}}\ and\ \bibinfo {author} {\bibfnamefont {A.~A.}\ \bibnamefont {Migdal}},\ }\href {\doibase 10.1016/0550-3213(88)90146-0} {\bibfield  {journal} {\bibinfo  {journal} {Nucl. Phys.}\ }\textbf {\bibinfo {volume} {B311}},\ \bibinfo {pages} {171} (\bibinfo {year} {1988})}\BibitemShut {NoStop}%
\bibitem [{\citenamefont {Brezin}\ and\ \citenamefont {Kazakov}(1990)}]{Brezin:1990rb}%
  \BibitemOpen
  \bibfield  {author} {\bibinfo {author} {\bibfnamefont {E.}~\bibnamefont {Brezin}}\ and\ \bibinfo {author} {\bibfnamefont {V.~A.}\ \bibnamefont {Kazakov}},\ }\href {\doibase 10.1016/0370-2693(90)90818-Q} {\bibfield  {journal} {\bibinfo  {journal} {Phys. Lett.}\ }\textbf {\bibinfo {volume} {B236}},\ \bibinfo {pages} {144} (\bibinfo {year} {1990})}\BibitemShut {NoStop}%
\bibitem [{\citenamefont {Gross}\ and\ \citenamefont {Migdal}(1990)}]{Gross:1989vs}%
  \BibitemOpen
  \bibfield  {author} {\bibinfo {author} {\bibfnamefont {D.~J.}\ \bibnamefont {Gross}}\ and\ \bibinfo {author} {\bibfnamefont {A.~A.}\ \bibnamefont {Migdal}},\ }\href {\doibase 10.1103/PhysRevLett.64.127} {\bibfield  {journal} {\bibinfo  {journal} {Phys. Rev. Lett.}\ }\textbf {\bibinfo {volume} {64}},\ \bibinfo {pages} {127} (\bibinfo {year} {1990})},\ \bibinfo {note} {[,127(1989)]}\BibitemShut {NoStop}%
\bibitem [{\citenamefont {Parisi}(1990)}]{Parisi:1989dka}%
  \BibitemOpen
  \bibfield  {author} {\bibinfo {author} {\bibfnamefont {G.}~\bibnamefont {Parisi}},\ }\href {\doibase 10.1016/0370-2693(90)91722-N} {\bibfield  {journal} {\bibinfo  {journal} {Phys. Lett.}\ }\textbf {\bibinfo {volume} {B238}},\ \bibinfo {pages} {209} (\bibinfo {year} {1990})}\BibitemShut {NoStop}%
\bibitem [{\citenamefont {Ginsparg}\ and\ \citenamefont {Zinn-Justin}(1990)}]{Ginsparg:1990as}%
  \BibitemOpen
  \bibfield  {author} {\bibinfo {author} {\bibfnamefont {P.~H.}\ \bibnamefont {Ginsparg}}\ and\ \bibinfo {author} {\bibfnamefont {J.}~\bibnamefont {Zinn-Justin}},\ }\href {\doibase 10.1016/0370-2693(90)91108-N} {\bibfield  {journal} {\bibinfo  {journal} {Phys. Lett.}\ }\textbf {\bibinfo {volume} {B240}},\ \bibinfo {pages} {333} (\bibinfo {year} {1990})}\BibitemShut {NoStop}%
\bibitem [{\citenamefont {Das}\ and\ \citenamefont {Jevicki}(1990)}]{Das:1990kaa}%
  \BibitemOpen
  \bibfield  {author} {\bibinfo {author} {\bibfnamefont {S.~R.}\ \bibnamefont {Das}}\ and\ \bibinfo {author} {\bibfnamefont {A.}~\bibnamefont {Jevicki}},\ }\href {\doibase 10.1142/S0217732390001888} {\bibfield  {journal} {\bibinfo  {journal} {Mod. Phys. Lett.}\ }\textbf {\bibinfo {volume} {A5}},\ \bibinfo {pages} {1639} (\bibinfo {year} {1990})},\ \bibinfo {note} {[,1639(1990)]}\BibitemShut {NoStop}%
\bibitem [{\citenamefont {Polchinski}(1991)}]{Polchinski:1991uq}%
  \BibitemOpen
  \bibfield  {author} {\bibinfo {author} {\bibfnamefont {J.}~\bibnamefont {Polchinski}},\ }\href {\doibase 10.1016/0550-3213(91)90559-G} {\bibfield  {journal} {\bibinfo  {journal} {Nucl. Phys. B}\ }\textbf {\bibinfo {volume} {362}},\ \bibinfo {pages} {125} (\bibinfo {year} {1991})}\BibitemShut {NoStop}%
\bibitem [{\citenamefont {Boulatov}\ and\ \citenamefont {Kazakov}(1993)}]{Boulatov:1991xz}%
  \BibitemOpen
  \bibfield  {author} {\bibinfo {author} {\bibfnamefont {D.}~\bibnamefont {Boulatov}}\ and\ \bibinfo {author} {\bibfnamefont {V.}~\bibnamefont {Kazakov}},\ }\href {\doibase 10.1142/S0217751X9300031X} {\bibfield  {journal} {\bibinfo  {journal} {Int. J. Mod. Phys. A}\ }\textbf {\bibinfo {volume} {8}},\ \bibinfo {pages} {809} (\bibinfo {year} {1993})},\ \Eprint {http://arxiv.org/abs/hep-th/0012228} {arXiv:hep-th/0012228} \BibitemShut {NoStop}%
\bibitem [{\citenamefont {Gross}\ and\ \citenamefont {Klebanov}(1991)}]{Gross:1990md}%
  \BibitemOpen
  \bibfield  {author} {\bibinfo {author} {\bibfnamefont {D.~J.}\ \bibnamefont {Gross}}\ and\ \bibinfo {author} {\bibfnamefont {I.~R.}\ \bibnamefont {Klebanov}},\ }\href {\doibase 10.1016/0550-3213(91)90363-3} {\bibfield  {journal} {\bibinfo  {journal} {Nucl. Phys. B}\ }\textbf {\bibinfo {volume} {354}},\ \bibinfo {pages} {459} (\bibinfo {year} {1991})}\BibitemShut {NoStop}%
\bibitem [{\citenamefont {Gross}\ and\ \citenamefont {Klebanov}(1990)}]{Gross:1990ub}%
  \BibitemOpen
  \bibfield  {author} {\bibinfo {author} {\bibfnamefont {D.~J.}\ \bibnamefont {Gross}}\ and\ \bibinfo {author} {\bibfnamefont {I.~R.}\ \bibnamefont {Klebanov}},\ }\href {\doibase 10.1016/0550-3213(90)90667-3} {\bibfield  {journal} {\bibinfo  {journal} {Nucl. Phys. B}\ }\textbf {\bibinfo {volume} {344}},\ \bibinfo {pages} {475} (\bibinfo {year} {1990})}\BibitemShut {NoStop}%
\bibitem [{\citenamefont {Kazakov}\ \emph {et~al.}(2002)\citenamefont {Kazakov}, \citenamefont {Kostov},\ and\ \citenamefont {Kutasov}}]{Kazakov:2000pm}%
  \BibitemOpen
  \bibfield  {author} {\bibinfo {author} {\bibfnamefont {V.}~\bibnamefont {Kazakov}}, \bibinfo {author} {\bibfnamefont {I.~K.}\ \bibnamefont {Kostov}}, \ and\ \bibinfo {author} {\bibfnamefont {D.}~\bibnamefont {Kutasov}},\ }\href {\doibase 10.1016/S0550-3213(01)00606-X} {\bibfield  {journal} {\bibinfo  {journal} {Nucl. Phys. B}\ }\textbf {\bibinfo {volume} {622}},\ \bibinfo {pages} {141} (\bibinfo {year} {2002})},\ \Eprint {http://arxiv.org/abs/hep-th/0101011} {arXiv:hep-th/0101011} \BibitemShut {NoStop}%
\bibitem [{\citenamefont {Alexandrov}\ \emph {et~al.}(2002)\citenamefont {Alexandrov}, \citenamefont {Kazakov},\ and\ \citenamefont {Kostov}}]{Alexandrov:2002fh}%
  \BibitemOpen
  \bibfield  {author} {\bibinfo {author} {\bibfnamefont {S.~Y.}\ \bibnamefont {Alexandrov}}, \bibinfo {author} {\bibfnamefont {V.~A.}\ \bibnamefont {Kazakov}}, \ and\ \bibinfo {author} {\bibfnamefont {I.~K.}\ \bibnamefont {Kostov}},\ }\href {\doibase 10.1016/S0550-3213(02)00541-2} {\bibfield  {journal} {\bibinfo  {journal} {Nucl. Phys. B}\ }\textbf {\bibinfo {volume} {640}},\ \bibinfo {pages} {119} (\bibinfo {year} {2002})},\ \Eprint {http://arxiv.org/abs/hep-th/0205079} {arXiv:hep-th/0205079} \BibitemShut {NoStop}%
\bibitem [{\citenamefont {Gromov}\ \emph {et~al.}(2014)\citenamefont {Gromov}, \citenamefont {Kazakov}, \citenamefont {Leurent},\ and\ \citenamefont {Volin}}]{Gromov:2013pga}%
  \BibitemOpen
  \bibfield  {author} {\bibinfo {author} {\bibfnamefont {N.}~\bibnamefont {Gromov}}, \bibinfo {author} {\bibfnamefont {V.}~\bibnamefont {Kazakov}}, \bibinfo {author} {\bibfnamefont {S.}~\bibnamefont {Leurent}}, \ and\ \bibinfo {author} {\bibfnamefont {D.}~\bibnamefont {Volin}},\ }\href {\doibase 10.1103/PhysRevLett.112.011602} {\bibfield  {journal} {\bibinfo  {journal} {Phys. Rev. Lett.}\ }\textbf {\bibinfo {volume} {112}},\ \bibinfo {pages} {011602} (\bibinfo {year} {2014})},\ \Eprint {http://arxiv.org/abs/1305.1939} {arXiv:1305.1939 [hep-th]} \BibitemShut {NoStop}%
\bibitem [{\citenamefont {Gromov}\ \emph {et~al.}(2015)\citenamefont {Gromov}, \citenamefont {Kazakov}, \citenamefont {Leurent},\ and\ \citenamefont {Volin}}]{Gromov:2014caa}%
  \BibitemOpen
  \bibfield  {author} {\bibinfo {author} {\bibfnamefont {N.}~\bibnamefont {Gromov}}, \bibinfo {author} {\bibfnamefont {V.}~\bibnamefont {Kazakov}}, \bibinfo {author} {\bibfnamefont {S.}~\bibnamefont {Leurent}}, \ and\ \bibinfo {author} {\bibfnamefont {D.}~\bibnamefont {Volin}},\ }\href {\doibase 10.1007/JHEP09(2015)187} {\bibfield  {journal} {\bibinfo  {journal} {JHEP}\ }\textbf {\bibinfo {volume} {09}},\ \bibinfo {pages} {187} (\bibinfo {year} {2015})},\ \Eprint {http://arxiv.org/abs/1405.4857} {arXiv:1405.4857 [hep-th]} \BibitemShut {NoStop}%
\bibitem [{Note2()}]{Note2}%
  \BibitemOpen
  \bibinfo {note} {The original expression in \cite {Balog:1992cm} was computed in the $\protect \overline {MS}$ scheme. A finite renormalization constant connecting $\protect \overline {MS}$ to the scheme defined by (\ref {running-lambda}) was computed in \cite {Fateev:1994ai}, and has been already taken into account in (\ref {E1-loop}).}\BibitemShut {Stop}%
\bibitem [{\citenamefont {Mari\~no}\ and\ \citenamefont {Reis}(2020)}]{Marino:2019eym}%
  \BibitemOpen
  \bibfield  {author} {\bibinfo {author} {\bibfnamefont {M.}~\bibnamefont {Mari\~no}}\ and\ \bibinfo {author} {\bibfnamefont {T.}~\bibnamefont {Reis}},\ }\href {\doibase 10.1007/JHEP04(2020)160} {\bibfield  {journal} {\bibinfo  {journal} {JHEP}\ }\textbf {\bibinfo {volume} {04}},\ \bibinfo {pages} {160} (\bibinfo {year} {2020})},\ \Eprint {http://arxiv.org/abs/1909.12134} {arXiv:1909.12134 [hep-th]} \BibitemShut {NoStop}%
\bibitem [{\citenamefont {Hartnoll}\ and\ \citenamefont {Mazenc}(2015)}]{Hartnoll:2015fca}%
  \BibitemOpen
  \bibfield  {author} {\bibinfo {author} {\bibfnamefont {S.~A.}\ \bibnamefont {Hartnoll}}\ and\ \bibinfo {author} {\bibfnamefont {E.}~\bibnamefont {Mazenc}},\ }\href {\doibase 10.1103/PhysRevLett.115.121602} {\bibfield  {journal} {\bibinfo  {journal} {Phys. Rev. Lett.}\ }\textbf {\bibinfo {volume} {115}},\ \bibinfo {pages} {121602} (\bibinfo {year} {2015})},\ \Eprint {http://arxiv.org/abs/1504.07985} {arXiv:1504.07985 [hep-th]} \BibitemShut {NoStop}%
\bibitem [{\citenamefont {V.Kazakov}\ \emph {et~al.}()\citenamefont {V.Kazakov}, \citenamefont {E.Sobko},\ and\ \citenamefont {K.Zarembo}}]{KSZ_Long}%
  \BibitemOpen
  \bibfield  {author} {\bibinfo {author} {\bibnamefont {V.Kazakov}}, \bibinfo {author} {\bibnamefont {E.Sobko}}, \ and\ \bibinfo {author} {\bibnamefont {K.Zarembo}},\ }\href@noop {} {\ }\Eprint {http://arxiv.org/abs/in preparation} {in preparation} \BibitemShut {NoStop}%
\bibitem [{\citenamefont {Di~Pietro}\ \emph {et~al.}(2021)\citenamefont {Di~Pietro}, \citenamefont {Mari\~no}, \citenamefont {Sberveglieri},\ and\ \citenamefont {Serone}}]{DiPietro:2021yxb}%
  \BibitemOpen
  \bibfield  {author} {\bibinfo {author} {\bibfnamefont {L.}~\bibnamefont {Di~Pietro}}, \bibinfo {author} {\bibfnamefont {M.}~\bibnamefont {Mari\~no}}, \bibinfo {author} {\bibfnamefont {G.}~\bibnamefont {Sberveglieri}}, \ and\ \bibinfo {author} {\bibfnamefont {M.}~\bibnamefont {Serone}},\ }\href {\doibase 10.1007/JHEP10(2021)166} {\bibfield  {journal} {\bibinfo  {journal} {JHEP}\ }\textbf {\bibinfo {volume} {10}},\ \bibinfo {pages} {166} (\bibinfo {year} {2021})},\ \Eprint {http://arxiv.org/abs/2108.02647} {arXiv:2108.02647 [hep-th]} \BibitemShut {NoStop}%
\end{thebibliography}%

\end{document}